\documentclass[12pt]{iopart}
\usepackage{iopams}

\usepackage[dvips]{graphicx}
\usepackage[dvips]{color}
\DeclareGraphicsExtensions{.eps}

\newcommand{\BE}{\begin{equation}}
\newcommand{\EE}{\end{equation}}
\newcommand{\BA}{\begin{eqnarray}}
\newcommand{\EA}{\end{eqnarray}}
\begin{document} 

\title{Linear frictional forces cause orbits to neither 
circularize nor precess} 

\author{B. Hamilton$^{1,2}$ and M. Crescimanno$^1$} 
\address{$^1$ Dept. of Physics and Astronomy, Youngstown State University}
\address{$^2$ Dept. of Physics, University of Maryland}
\ead{\mailto{bkham@mail.umd.edu, mcrescim@cc.ysu.edu}}


\submitto{\JPA}
\eqnobysec

\vskip .1in 

\maketitle 

\input epsf.tex

\noindent {\bf Abstract:} 
For the undamped Kepler potential the 
lack of precession has historically been understood in terms of
the Runge-Lenz symmetry. For the damped Kepler problem 
this result may be understood in terms
of the generalization of Poisson structure to damped systems
suggested recently by Tarasov\cite{Tarasov}. In this generalized algebraic
structure the orbit-averaged Runge-Lenz vector remains a constant 
in the linearly damped Kepler problem to leading 
order in the damping coefficient. Beyond Kepler, we prove that,
for any potential  proportional to a power of the radius,  the 
orbit shape and precession angle remain constant to leading order in 
the linear friction coefficient. 

\vskip .3in 
\section{Introduction}  
What happens to orbits subject to linear frictional drag? 
In typical physical settings, such as Rydberg 
atoms or stellar binaries, the effective frictional 
forces are nonlinear and, typically, lead to 
the circularization of the orbit. Orbital evolution under
linear friction is special in that, as we show below, 
the eccentricity and the apsides do not change to leading order
in the damping. The purpose of this note 
is to understand this elementary result from the underlying 
dynamical symmetry of the Kepler problem, 
thus demonstrating the utility of a 
Hamiltonian notion in 
its non-Hamiltonian generalization.

In many astrophysical 
situations, the secular evolution due to friction of orbits in a two body 
system is towards circular orbits. 
In an orbit in a central field the 
angular momentum scales with the momentum while the energy generally scales 
with the
momentum-squared. Friction, assumed to be spatially isotropic and 
homogeneous but time 
odd, typically scales the momentum. This means generally that the resultant 
secular evolution in central force systems is that in which the energy is 
minimized 
at fixed angular momentum. 
This is clearly the circular orbit. The velocity dependence of the 
frictional force is quite relevant, in particular as referenced against the 
velocity dispersion of the (undamped) motion in that central potential.
Clearly, under the action of such dissipative forces, a consequence 
of symmetry is that the flow in orbital shape (not size!)
has two fixed points, circular orbits and 
strictly radial (infall) orbits.

Few physical problems have received 
more scrutiny than
bounded orbits in the two- and few- body system. 
Among these, the two-body Kepler problem 
is arguably the most experimentally relevant and best 
studied example, having  
been illuminated by intense theoretical inquiry spanning
hundreds of years leading to 
important insights even in relatively recent times
\cite{Ermano, Hermann, Bernoulli, Laplace, Runge, Lenz, Pauli, Hulthen, Fock, Bargmann, Moser, Belbruno, Ligon, Cushman, Stehle, Iosifescu, Gyorgyi, Guillemin, Cordani, Keane1, Fradkin}. 
We do not present a systematic review or histiography of this 
celebrated problem (though we thankfully acknowledge also 
\cite{Keane2, Goldstein, Jose, Quesne, Kustaanheimo, Chen, 
Stahlhofen, Landau1} 
which we have found quite useful for our study). 
We do not aim to contribute 
to the vast literature on astrophysically and microphysically 
relevant models of friction in orbital problems (though the interested 
reader may find references
\cite{Colosimo, Casertano, Mayor, Mardling, Kozai1, Kozai2, Apostolatos, Dotti, Glampedakis, Apostalakis, Lightman, Schutz,  Bellomo} 
a useful launching point for such review). 

Instead, the our purpose here is to accomodate from the 
dynamical symmetry group point-of-view the result that 
linear frictional damping (to leading order) preserves 
the orbit's shape.  
Although Hamiltonian systems may lose dynamical symmetry
completely when dissipative forces are included, it can be shown
that some structure may remain under a modified symplectic form. 
After a brief introduction to the method by which Tarasov
extends symplectic structure of Hamiltonian mechanics to 
dissipative systems, we apply it to the determination of the 
time averages of dynamical quantities. Damping 
invariably introduces new dynamical timescales and the 
time averaging we implement is over times short compared
with these timescales (but still long compared with the 
orbital timescales in the undamped problem). 
Tarasov's construction 
reveals the relevance of the dynamical symmetry algebra 
to the damped Kepler problem. 

We then compare this aproach to 
the classic ``variations of constants'' method of 
orbit parameter evolution 
by describing an
improvement 
that follows from our study. The elementary method 
can be generalized to non-Kepler homogeneous potentials and also  
determines orbital shape evolution for linearly 
damped Kepler orbits beyond leading order. 

\section{Dynamical Symmetry and Tarasov's Construction} 
In the undamped Kepler problem the lack of precession is generally
understood as a consequence of a dynamical symmetry, the 
celebrated $so(4)$ symmetry formed from the two commuting $so(3)$, 
one from the angular momentum ${\vec L} = {\vec r}\times{\vec p}$
 the other from the Runge-Lenz vector,
${\vec S} = {\vec L}\times{\vec p} + k{{\vec r}\over{|{\vec r}|}}$
(\cite{Runge, Lenz, Pauli})
being the maximal set of local, algebraically independent
operators that commute with the Hamiltonian, $H={{ {\vec p}^2}\over{2}} + V(r) $, with $V(r) = kr^\alpha$ for $k < 0$ and $\alpha = -1$. 
(though see \cite{Fradkin} for a more precise and general statement of the 
connection between algebra and orbits in a central field). 
\BE 
\fl 
\{L_i, L_j\} = 2\epsilon_{ijk} L_k \qquad \{L_i, S_j\} = 2\epsilon_{ijk} S_k \qquad 
\{S_i, S_j\} = -2H\epsilon_{ijk} S_k
\label{14}
\EE
The length of ${\vec S}$ is 
proportional to the eccentricity (and points along the 
semi-major axis of the orbit, in the
direction to the periastron from the focus). 
Defining ${\vec L}$ and ${\vec S}$ has utility beyond their being 
constants in the 2-body
Kepler problem, for example, parameterizing the secular evolution of orbits 
under various 
Hamiltonian perturbations \cite{Stahlhofen, Morehead}. 
This  $so(4)$ is one of the maximal compact factor groups of the 
$so(4,2)$ (the conformal group) 
extended symmetry formed by ${\vec L}, {\vec A}, H$, 
the generalization of the scaling operator
${\cal R} = {\vec r}\cdot{\vec p}$ and the 
Virial operator 
${\cal V} ={{ {\vec p}^2}\over{2}} - {{r}\over{2}}\partial_r V(r) $ 
(\cite{Iosifescu, Cordani}) 

The other central potential posessing an easily recognizable 
dynamical symmetry is the
multi-dimensional harmonic oscillator ($V$ as given with $k > 0$, 
$\alpha$ = 2). As is well known, the 
isotropic $D$-dimensional
harmonic oscillator's naive $O(D)$ symmetry is part of a larger $U(D)$ 
dynamical symmetry. For $D=2$ harmonic oscillator, note that the 
$U(2)$ symmetry does enlarge further to a $so(3,2)$ when including  
${\cal R}, {\cal V}$ and their generalization (the virial subalgebra \Eref{4} through 
\Eref{6} of each oscillator alone and closes 
to a $sl(2,R)$ subgroup of the $so(3,2)$). Note 
further that it is this later algebra that is isomorphic to the 
dimensionally reduced 
$so(4,2)$ of the 3-d Kepler problem, by which we mean the reduction of that algebra 
to generators associated with the orbital plane only. These considerations 
can also be understood from the KS construction\cite{Kustaanheimo, Chen}
of the Kepler problem, in which a four-dimensional 
isotropic harmonic oscillator is the starting point. 
In that construction the 
$u(4)  = su(4) ~{\times}~ u(1)$ is, of itself, not preserved by the KS construction. 
Instead, it is the $u(2,2)$ subgroup of the four 
identical, independent
oscillator's 
$sp(8,{\bf {\rm R}})$ symmetry in which the overall $u(1)$ can be isolated as
the angular momentum constraint of the KS construction\cite{Quesne}. 
The residual symmetry $su(2,2) \sim so(4,2)$ is 
that of the 3-d Kepler problem. 
The analytical connection between the Kepler problem and the 
isotropic harmonic oscillator has deep historical roots, going back to 
Newton and Hooke (see \cite{Grant} and references therein). 
Finally, the geometric 
construction of the undamped Kepler problem as geodesic flow on (spatial) 
a 3-manifolds of constant curvature relates the $so(4)$ dynamical 
symmetry to the isometry group generated by Killing 
vector fields on the spatial slice\cite{Guillemin, Keane1, Keane2}. 
These various connections between the Kepler problem and 
the isotropic harmonic oscillator do not lead to a 
simple structural connection between the associated damped problems.

To leading order in the damping, Kepler 
orbits subject to linear frictional force 
do not change shape or precess as they decay. 
It would be satisfying to understand this elementary 
result as a consequence of the preservation of 
the dynamical algebra under linear friction. Although this is 
reminiscent of the damped N-dimensional harmonic oscillator, 
there is no simple way to relate the damped problems. 
Since the subgroup 
associated with the shape and precession (through the ${\vec S}$) is rank one
it is suggestive that the entire group structure is preserved 
to leading order in the linear friction. 

A recent paper by Tarasov\cite{Tarasov} 
suggests a straightforward generalization 
of the Poisson structure to systems with dissipative forces. 
There are many other approaches to addressing 
structural questions of dissipative systems
(for one example, see \cite{Sergi1, Sergi2}). We find the approach of
\cite{Tarasov} to be most useful for addressing questions of the 
dynamical symmetries that survive including dissipation. 
For completeness we now briefly review Tarasov's construction, and 
apply it to dissipation in the central field problem 
in the following section. 

To preserve as much of the algebraic structure as 
possible, Tarasov constructs a one-parameter
family of two forms (that define a generalized Poisson structure) 
that -in a sense- 
interpolate between different dampings.
In the zero damping limit it smoothly matches onto the 
canonical symplectic form. 
Dimensionally, any damping parameter introduces a new time
scale into the problem, thus this new interpolating two form 
must also be explicitely time-dependent.  
Tarasov requires this family of two forms to 
have the following useful properties

(1) Non-degeneracy: The two form 
${\bf \omega} = \omega_{ij}(t) {\bf {\rm d}}x^i\wedge{\bf {\rm d}}x^j$ 
is antisymmetric and non-degenerate along the entire flow. The $x^i$ are the 
$2N$ (local) phase space co-ordinates. In positive terms, the
inverse $\omega^{ij}\omega_{jk} = \delta^i_j$ exists almost 
globally \footnote{Since we do not formulate this entirely in the 
exterior calculus, we must allow for higher codimension singularites 
that may not be resolvable in the dissipative system.}. 

(2) Jacobi Identity: the two form is used to define 
a new Poisson bracket 
$\{A, B\}_T = \omega^{ij}\partial_i A \partial_j B$ that 
forms an associative algebra. Explicitely it satisfies. 
\BE 
\{A,\{B,C\}_T\}_T + \{B,\{C,A\}_T\}_T +\{C,\{A,B\}_T\}_T =0
\label{jacobi}
\EE
Here we use the subscript '$T$' to distinguish this bracket
from the Poisson bracket of the undamped problem. 

(3) Derivation property of time translation: 
with respect to this new
bracket the time derivative of the new
Poisson bracket satisfies the derivation 
property (also called the Liebnitz rule) 
\BE
{{\rm d}\over{\rm d}t} \{A, B\}_T = \{ { {{\rm d}A}\over{\rm d}t}, B\}_T + 
\{A, {{{\rm d}B}\over{\rm d}t} \}_T
\label{derivation} 
\EE

These requirements are remarkable for several reasons. 
First, property (1) indicates that (2) and (3) are possible.
The deeper relevance of property (1) is that we 
can regard the two-form as (essentially a) global metric 
on the phase space. Property (2) indicates local mechanical 
observables in 
this 'dissipation deformed' algebra form a lie algebra. 
Property (3) is key to the utility of Tarasov's construction for 
understanding constants of motion in dissipative systems. 
It stipulates that time development 
in the dissipative system, while no longer just $\{~, H\}$ (or even 
$\{~, H\}_T$), must 
be compatible with the 
structure of the symplectic algebra in the new bracket
and thus the (new) bracket of 
time independent quantities in the dissipative system are themselves
time independent. Thus, just as in the Hamiltonian case, 
time independent quantities form a closed subalgebra. 
Note that for a Hamiltonian system property 
(3) is automatic since in that case time translation is an 
inner automorphism of the symplectic algebra. 
In a dissipative system by contrast
the Hamiltonian is no longer the operator of time translation, but,  
if Tarasov's construction can be implemented, time translation is still 
an automorphism of the algebra, and as such may be regarded 
as an outer automorphism. 
Finally, from property (3) it follows after a brief calculation  
that the two-form ${\bf \omega}$ must be time idependent in the full 
dissipative system, ${ {{\rm d}{\bf \omega}}\over{\rm d}t} =0$. In 
terms of symplectic geometry, this is metric compatibility of the 
dissipative flow. 

To proceed with the construction, consider the general 
flow ${\dot x}^i = \chi^i({\vec x}, t)$. Again, these are not assumed to 
be Hamiltonian flows. 
Assuming property (1) and using $\omega^{ij}$ to form a bracket
$\{A,B\}_T = \omega^{ij} \partial_i A \partial_j B$ , 
property (2) leads to the condition
\BE 
\omega^{im}\partial_{m}\omega^{jk} + \omega^{jm}\partial_{m}\omega^{ki} 
+ \omega^{km}\partial_{m}\omega^{ij}  = 0 . 
\label{jacobi2} 
\EE
Total time derivatives and derivatives along phase space
directions do not commute in the flow, 
\BE 
[{{ {\rm d}}\over{{\rm d} t}}, \partial_i ] A = -\partial_j A \partial_i \chi^j . 
\label{time}
\EE
Using this and the jacobi identity \eref{jacobi}, one sees that
property (3) implies a condition relating the form $\bf \omega$ 
and the flow $\chi^i$, 
\BE 
{{\partial \omega_{ij}(t)}\over{\partial t}} = \partial_i \chi_j - \partial_j \chi_i 
\qquad {\rm where} \qquad \chi_j = \omega_{jk}(t) \chi^k
\label{derivation_explicit}
\EE
Given $\chi^i$, we proceed by solving 
\eref{jacobi2} for an $\omega^{ij}$ 
that satisfies \eref{derivation_explicit}. This 
completes Tarasov's construction. 

We breifly offer a few further remarks 
helpful to orient the reader.
First, in the more familiar context of Hamiltonian flows, there
${\dot x}^i = \chi^i = \{x^i, H\}$ for a local function $H$ on the 
phase space. For this case we can compute in the Darboux 
frame and learn that the usual symplectic form 
(automatically satisfying \eref{jacobi2})
is a solution also to \eref{derivation_explicit} since the RHS in that 
case is zero. We recognize the RHS of \eref{derivation_explicit} as
exactly the obstruction to the flow, $\chi^i$ being Hamiltonian. 

Conformal transformation 
of the two-form, $\tilde {\bf \omega} = \Omega {\bf \omega}$, 
where $\Omega$ is a a scalar function, can only relate two solutions
of \eref{jacobi2} and \eref{derivation_explicit} IFF the $\Omega$
is a constant of the motion ${ {{\rm d}\Omega}\over{{\rm d}t}} = 0$. 
For in that case \eref{derivation_explicit} indicates that 
\BE
{{\partial \Omega}\over{ \partial t}} \omega_{ij} = 
\chi_j\partial_i\Omega - \chi_i\partial_j\Omega
\label{timeprime}
\EE 
whereas \eref{jacobi2} yields
\BE 
\omega_{jk}\partial_j\Omega + \omega_{kl}\partial_j\Omega + 
\omega_{lj}\partial_k\Omega = 0
\label{derivation_explicitprime}
\EE
so, contracting by $\chi^k$ and comparing with \eref{timeprime}, we
learn that $\Omega$ must be a constant of the motion. Thus, each
solution is conformally unique.  

We do not know what conditions on $\chi^i$ 
lead to the existence of even one non-singular 
{\it simultaneous} solution ${\bf \omega}$
of \eref{jacobi2}
and 
\eref{derivation_explicit}. 
Tarasov\cite{Tarasov} provides an explicit solution
for a general Hamiltonian system ammended by a general 
linear frictional force.
The general question of the 
existence of ${\bf \omega}(t)$ for a more general $\chi^i$ is at this 
point unclear, but beyond the scope of this present effort.

\section{Dynamical Symmetry in a Damped System} 
Consider damped orbital motion in a central field; 
\BE
 {\dot {\vec x}} = {\vec p}
\label{7}
\EE

\BE 
{\dot {\vec p}} = -\partial_r V {{\vec x}\over{r}} 
-\beta(p) {\vec p}
\label{8}
\EE
with $r=|{\vec x}|$ and 
$V(r)$ the interparticle potential (throughout we take the reduced 
mass to be normalized to 1). The function $\beta(p)$ is some general 
function parameterizing the speed dependence of the damping, and 
this form of the damping function is the most general consistent 
with isotropy and homogeniety of the damping forces. 
Note that we can understand this set as descending from a limit 
in which the central mass is very much larger than the orbital mass though,
as in general,
damping does inextricably mix the center of mass motion and the 
relative motion. 
We call linear damping the choice of 
$\beta$ constant. 

The 
\Eref{derivation_explicit} takes the form, 
\BE  {{\partial \omega_{xp}(t)}\over{\partial t}} = 
\partial_x (\omega_{px} \chi^x) - \partial_p (\omega_{xp} \chi^p) 
 = \partial_p \omega_{xp'} (\beta(p) p') 
\label{20}
\EE
Again, we do not know
if solutions to 
\Eref{20} exist and satisfy Jacobi for every choice of  $\beta(p)$. 
However, for
$\beta(p)=const.$ there is a 
simple solution to \Eref{20} that satisfies Jacobi\cite{Tarasov}, 
\BE  \omega_{ij}(t) = e^{\beta t}{\hat \omega}_{ij} 
\label{21}
\EE
where $\hat \omega$ is the usual symplectic form of the undamped  
Kepler problem. Physically this 
corresponds to the uniform shrinkage of phase space volumes 
under linear damping. 

Clearly, in going from 
$\{,\}$ (Poisson bracket) to the new bracket $\{,\}_T$ the relations in 
\Eref{14} gain a factor of $e^{-\beta t}$. The algebra 
in the new bracket resulting from 
this simple rescaling is still $so(4)$. The utility of this simple change 
to the algebra of \Eref{14} (which was for the undamped system) is that 
it is now compatible 
with the evolution under \Eref{7} and \Eref{8} of the damped system. To see this in 
an example, take the first relation in \Eref{14} and take the (total) time derivative 
of both sides. Then note 
${{ {\rm d} \{L_i,L_j\} }\over{{\rm d}t}} = -2\beta (2\epsilon_{ijk} L_k) \ne 
2\epsilon_{ijk} {\dot L}_k$; {\it i.e.} the usual Poisson bracket is no longer
compatible with time evolution. Duplicating the previous line for
$\{L_i, L_j\}_T = 2e^{-\beta t}\epsilon_{ijk} L_k$ one learns that this is compatible 
with the flow \Eref{7} and \Eref{8}. Similarly, one may check that all the brackets in 
\Eref{14} (after replacing $\{, \}$ with $\{, \}_T$) are as well. 
Also note that  $\{L_i, H\}_T = 0 = \{S_i, H\}_T$, though since 
brackets with $H$ 
no longer delineate time evolution, these equations do not 
imply that 
$\vec L$ and $\vec S$ are constants of the motion in the dissipative system
(also clear from \Eref{15} below). 

The critique here is familiar to any attempt to reconcile 
symplectic structure and dissipation; fundamentally, \Eref{7} and \Eref{8} still treat 
$x$ and $p$ differently so that time evolution is no longer an element in the dynamical 
algebra of $\{,\}$ or $\{, \}_T$.

To relax the category of 'constants of the motion' 
sufficiently for dissipative systems, 
consider to what extent 
dynamical quantities 
averaged over some number of orbits change
on a longer time scale, {\it i.e.} on 
a timescale relevant to the dissipation (note $1/\beta(p)$ 
is essentially that timescale). 
Let $<>$ denote time averages over many orbits, ${\cal O}$ a 
classical observable, 
and suppose that $\omega$
is a
solution to  
\Eref{derivation_explicit} 
and the Jacobi identity for the system as in 
\Eref{7} and \Eref{8}. 
In general, 
\BE
<\{ {\cal O}, H\}_T> = < \omega^{xp}(t) \bigl({\dot x}\partial_x {\cal O} 
 -(-{\dot p} - \beta(p)p) \partial_p {\cal O} \bigr) >
\EE
\BE 
\qquad \qquad \qquad = < \omega(t) \biggl[ { {{\rm d}{\cal O}}\over{ {\rm d} t}} - 
{{\partial {\cal O}}\over{\partial t}}\biggr] + \omega^{xp}(t)\beta(p) p \partial_p {\cal O} >
\label{23}
\EE 
Note that sums are implied in the $x, p$ indices of the $\omega^{xp}(t)$, 
the new symplectic form.
Above we have used isotropy to rewrite the sum in the first term in terms of 
the (normalized) symplectic trace of $\omega^{xp}(t)$ which we denote simply as $\omega(t)$.
To show one intermediate step, integrating by parts and using \Eref{20} we arrive at
$$
<\{ {\cal O}, H\}_T> = {{1}\over{T}} \Delta(\omega_T {\cal O}) +< \omega^{x\alpha}\omega^{p\beta}
(\partial_\alpha \chi_\beta - \partial_\beta \chi_\alpha + \chi^l\partial_l \omega_{\alpha\beta}){\cal O} 
$$
\BE
\qquad \qquad \qquad 
+ \omega^{xp}\beta(p)p\partial_p {\cal O} - \omega {{\partial {\cal O}}\over{\partial t}}> 
\label{24}
\EE
Where $\Delta(G)$ refers simply to the overall change of the 
quantity $G$ over time $T$. 
Finally, using the \Eref{time} and the fact that $\omega^{xp}$ satisfies the Jacobi 
identity we reduce the above to 
\BE 
\fl <\{ {\cal O}, H\}_T>  = {{1}\over{T}} \Delta(\omega {\cal O}) 
+ < (\omega^{xp'}\partial_{p'}\chi^p 
- \omega^{px'}\partial_{x'}\chi^x){\cal O}  + \omega^{xp}\beta(p)p\partial_p{\cal O} -
\omega{{\partial {\cal O}}\over{\partial t}}> 
\label{25}
\EE
We now specialize to vector fields 
of the general form \Eref{7} and \Eref{8} to find, 
\BE 
\fl {{1}\over{T}} \Delta(\omega {\cal O}) = <\{ {\cal O}, H\}_T + \omega { {\partial {\cal O}}\over
{\partial t}} 
- \omega^{xp} \bigl(\partial_p(\beta(p)) p {\cal O}-\beta(p) p \partial_p
{\cal O}\bigr) > 
\label{26}
\EE
and so making the RHS zero indicates conserved quantities in the non-Hamiltonian 
system. Again, this last result 
was derived for general $\beta(p)$, 
which assumes only that the friction is
isotropic and homogeneous. 
In the linear friction case $\beta(p) = \beta = const$.
For that case, using ${\cal O } = L/\omega^2$ in the above equation 
implies that $L/\omega $
are constants of the motion in this system. 
Similarly, taking ${\cal O} = S/\omega$
indicates that $\Delta S$ is proportional to $(2\beta {\vec L}/\omega)\times <\omega {\vec p}> $
which, again, is zero to first order in $\beta$. 
This result then applied to the case of bounded Kepler orbits with linear damping 
indicates that the (orbit-averaged) Runge-Lenz vector, and thus the dynamical algebra of the
Kepler problem,  is conserved to leading order in the linear friction coefficient. 

In elementary terms, although 
angular momentum ${\vec L}$ and 
${\vec S}$ are constants in the Hamiltonian system for 
$V(r) \sim {{1}\over{r}}$
they evolve under linear damping of \eref{7}, \eref{8} as, 
\BE 
{\dot {\vec L}} = -\beta {\vec L} \qquad \qquad \qquad {\dot {\vec S}} = -2\beta {\vec L}\times{\vec p}. 
\label{15}
\EE
Note that in the weak damping limit, since ${\vec L}$ is conserved to 
${\cal O}(\beta^0)$, the 
second equation time averages to 
$-2<\beta {\vec p}>\times<{\vec L}>$. 
Thus, again we learn that if the damping were 
strictly linear ($\beta$ constant) 
then since $<{\vec p}> = 0$, the time average 
of $\dot {\vec S}$ is $0$, again indicating that the eccentricity 
vector would be conserved to leading order. 
Note also that it is straightforward to 
integrate the ${\vec L}$ equation explicitely, finding 
${\vec L} = {\vec L}_0e^{-\beta t}$ the initial condition ${\vec L}_0$
being identified now a conserved quantity of the dissipative 
system. We use these results in the next section 
of this paper to ammend the 'textbook'  orbital secular evolution 
equations.

\section{The Damped Kepler Problem}  

The previous section suggests that (linear-) damped bounded Kepler orbits 
shrink but retain their aspect ratio and do not precess to leading order 
in the damping. It is well known that 
superlinear damping does lead to circularization whereas 
sublinear damping leads to infall orbits in the Kepler case. 
So far this begs the questions of whether this generalizes to other 
central field problems, and, if so, then at what order in the linear
damping coefficient do orbits undergo shape and precessional change. 
In this section we 
address both questions, 
first describing a problem that arises 
using a time-honored
pertubative method for treating general perturbing forces 
in the Kepler problem, and second,  
generalize the result of the preceeding section 
to a broad class of central field potentials. 
We then 
establish in precise terms the fate of Kepler orbits under 
linear damping. 

Consider the usual 
secular orbital evolution method (called ``the variations of 
constants'') most common in literature on cellestial mechanics, 
for example, in \cite{Danby} (Chapter 11 Section 5, pg. 323, though
see also the treatments of non-linear friction
in \cite{Watson, MurrayDermott, BC}). 
In the ``variations of constants' method, orbital response 
to an applied force 
${\vec F} = R{\vec x} + N{\vec L} + B {\vec L}\times{\vec x}$, in 
the orbit's tilt $\Omega$, the orbital plane's axis, $i$, the 
eccentricity $\epsilon$, the angle of the ascending node $\omega$ the 
semi-major axis $a$ and the period $T = 2\pi/n$ (in their notation) 
evolve following\cite{Danby}, 
\BE 
{{{\rm d}\Omega}\over{{\rm d}t}} = {{nar}\over{\sqrt{1-\epsilon^2}}} N {{\sin{u}}\over{\sin{i}}}
\label{d1}
\EE

\BE 
{{{\rm d} i}\over{{\rm d}t}} =  {{nar}\over{\sqrt{1-\epsilon^2}}} N \cos{u}
\label{d2}
\EE

\BE 
{{{\rm d} \omega}\over{{\rm d}t}} = 
{{n a^2 \sqrt{1-\epsilon^2}}\over{\epsilon}}
\bigl[-R\cos \theta + B (1+{{r}\over{P}})\sin\theta \bigr] - \cos{i} {{ {\rm d}\Omega}\over{{\rm d} t}}
\label{d3}
\EE

\BE 
{{{\rm d} \epsilon}\over{{\rm d}t}} =  n a^2 \sqrt{1-\epsilon^2}
\bigl[R\sin \theta + B (\cos\theta + \cos E)\bigr]
\label{d4}
\EE

\BE 
{{\rm d}a\over{{\rm d}t}} = {2n a^2} \bigl[ R {{a\epsilon}\over{\sqrt{1-\epsilon^2}}}\sin \theta + B {{a^2}\over{r}} \sqrt{1-\epsilon^2}\bigr]
\label{d5}
\EE

and where
\BE 
{{\rm d}n\over{{\rm d}t}} = -{{3n}\over{2a}} {{{\rm d}a}\over{ {\rm d}t}}
\label{d6}
\EE
with $u=\theta+\omega$ and for the unperturbed Kepler orbit, 
$ {{P}\over{r}} = 1 + \epsilon \cos\theta$, $P$ is the latus rectum, and $E$ is the anomaly, 
{\it i.e.} $r = P(1-\epsilon\cos E)$. 
The central angle $\theta$ is found via the usual definition of angular 
momentum.  
When we specialize these
Kepler orbit evolution equations to the 
case of isotropic and homogeneous friction we learn that 
(see \cite{Danby}, Chapter 11, section 7 but using $\beta(p) p$ 
for $T$ in that reference)  
\BE
{{{\rm d} a}\over{ {\rm d}t}} = 2pa^2 \beta(p) p
\label{d7} 
\EE

\BE
{{{\rm d}\omega}\over{ {\rm d}t}} = {{2\sin\theta}\over{\epsilon}} \beta(p)
\label{d8} 
\EE
and 
\BE
{{{\rm d} \epsilon}\over{ {\rm d}t}} = 2(\cos\theta + \epsilon)\beta(p)
\label{d9} 
\EE
We can now specialize further to the marginal case, linear friction 
$\beta(p) = \beta = const$. To integrate
these equations, note $r^2 {{{\rm d}\theta}\over{{\rm d}t}} = L = 
L_0 e^{-\beta t}$ and, in terms of the force components, $N=0$, and 
$R = \beta(p) p \cos{\upsilon}$ and $B = \beta(p) p \sin{\upsilon}$
where $\upsilon$ is the angle between the radius vector and the 
tangent to the orbit. That angle can be written 
using the parameteric form of $r$ in terms of the 
constants of the orbit and the angle $\theta$, 
($\sin\upsilon = L/rp$ and $\cos\upsilon = \epsilon\sin\theta/Lp$)
resulting in a self-contained  
pair of ODE's in $\epsilon$, $\theta$ and $t$, 

\BE 
{{{\rm d}\theta}\over{{\rm d}t}} = {{e^{+3\beta t}}\over{L_0^3}}
(1+\epsilon\cos\theta)^2 
\label{d10}
\EE

\BE
{{{\rm d}\epsilon}\over{{\rm d}t}} = -2\beta(\cos\theta + \epsilon)
\label{d11}
\EE
If we integrate these to leading order in $\beta$ only (by, for example, 
using the first equation to eliminate the time derivative to leading order
in $\beta$) we do indeed find that the eccentricity is an orbit-averaged
constant of the
motion. 
But difficulty arises when we try to understand these equations 
beyond leading order in the damping, as a direct numerical integration 
of the equation set reveals (Figure 1). For a broad set of initial 
angles and small initial eccentricities, 
the $\epsilon$ passes through zero and 
goes negative. For comparison, the eccentricity (i.e. the 
square root of the length of the ${\vec S}$ vector) computed by numerical 
integration of the original equations of motion for precisely 
the same mechanical parameters and initial conditions is included on that 
figure. 

\begin{figure} 
\includegraphics{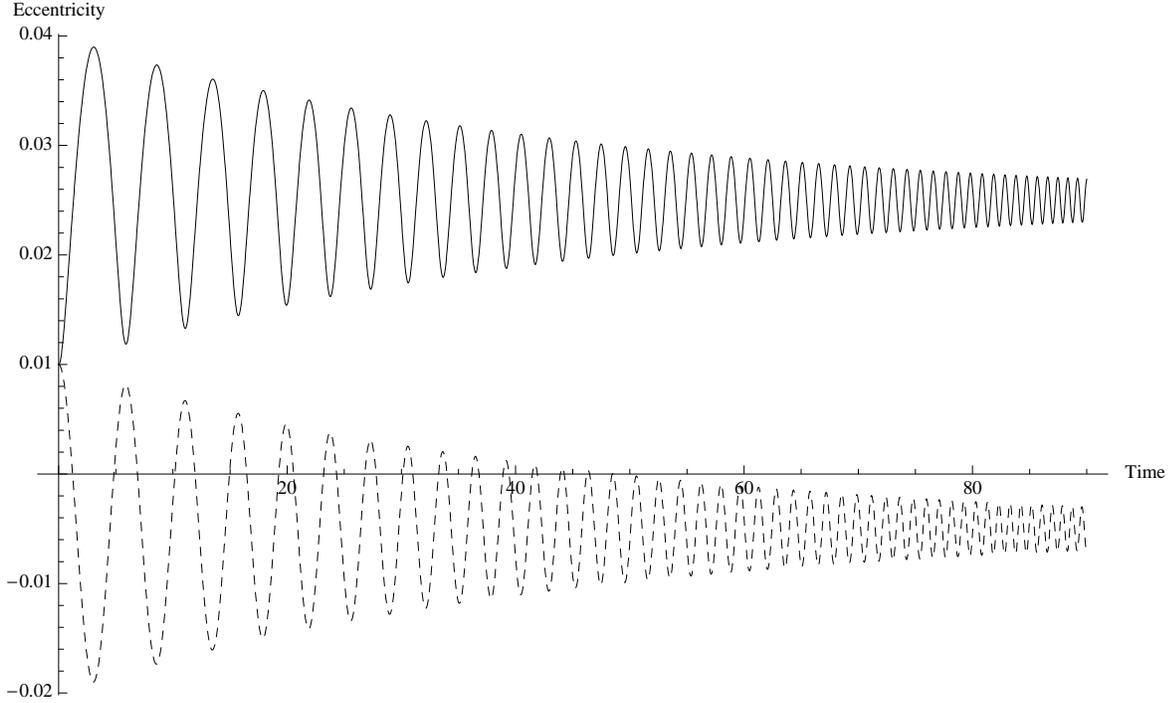}
\caption{The eccentricity in the actual damped Kepler problem (solid curve) compared with the eccentricity from \eref{d10} and \eref{d11} (dashed curve) versus time. Note 
that for the later the eccentricity can oscillates through zero and can even, as in this case, asymptote to a negative value.} 
\label{fig1} 
\end{figure} 

Even if one only wanted to assign importance to the 
asymptotic change in the eccentricity, that asymptotic change
from integrating the equation pair \eref{d10} , \eref{d11}
does not scale correctly with the
damping coefficient,  as may be checked numerically
(see \cite{Danby} for further admonisions against using the
``variations of constants'' method over long timescales). 
Clearly the ``variations of constants'' method at higher orders in the
evolution leads to unphysical results at short and long timescales. 
The fault is traceable to 
the fact that in higher order there are $\beta$-
(the damping coefficient) dependent terms in the orbit shape
whose
contributions are ignored substituting for $r$ 
using the undamped Kepler shape of the ellipse. 
This substitution is however
inectricably part of the 
``variation of constants'' method. 
To further clarify this problem with the 
``variations of constants'' method, 
it is not 
due to some ambiguity in the
eccentricity of a non-closed orbit, 
since eccentricity itself, rendered as the length of 
the ${\vec S}$ vector, has a local definition. Algebraically, 
with this definition of the eccentricity, 
note that $\epsilon^2 - 2L^2 U = k^2$ in the $1/r$ potential
{\it even under arbitrary damping}. A more useful 
algebraically identical form is 
$\epsilon^2 = 4{\cal V}^2r^2 -2 {\cal R}^2 H$, 
from which, since $H$ is negative for any damping 
function on a bounded orbit, we see immediately 
that $\epsilon^2$ is bounded away from zero. 

We now, in two parts, describe an approach emphasising the  
secular evolution of the dynamical symmetry, that addresses
this mismatch with the usual ``variation of constants'' method. 
For simplicity we focus in the main on 
potentials with fixed scaling wieght $\alpha$, deined through 
$ V(r) = kr^\alpha$. Orbits in any central potential are 
characterized by a fixed orbital plane and a single dimensionless
parameter, the ratio $d/c$ of the perihelion distance $d$ to the 
aphelion distance $c$. Let $L$ denote the angular momentum so that 
$ V_{eff}(r) = {{L^2}\over{2r^2}} + V(r)$ is the effective potential. 
Then from $V_{eff}(c) = U = V_{eff}(d)$ where $U$ is the total energy 
for a $V(r)$ of a fixed $\alpha$, we have, 
\BE
{{U}\over{k}} = {{c^{\alpha+2}-d^{\alpha+2}}\over{c^2-d^2}} \qquad \qquad
{{L^2}\over{2k}} = {{c^{\alpha}-d^{\alpha}}\over{c^2-d^2}} c^2 d^2
\label{9}
\EE
that then can be reduced to a 'dispersion relation' between $U$ and $L$,  
\BE 
{{U^{\alpha+2}}\over{k^2 L^{2\alpha}}} = f(d/c)
\label{10}
\EE
where $f$ in this case is a monotonic function on $[0,1]$.  Note also 
that $f(x) = f(1/x)$. We call $d/c$ the aspect ratio of the orbit 
(related to the eccentricity in the $\alpha = -1$ case). 
Thus in  leading order (only) in the damping we think of the
RHS as a function of the orbital eccentricity only. 
In applications, the differential form of \eref{10} is particularly useful, 
\BE 
\biggl[(\alpha+2) {{\delta U}\over{\delta L}} -2\alpha {{U}\over{L}} \biggr] {{\delta L}\over{U}}= 
{{f'}\over{f}} \delta (d/c)
\label{11}
\EE 
Equations of this sort are often written down when referring to the secular evolution 
of orbital system (see for example \cite{PJ} and references therein). 
As before consider further only damping forces that are isotropic 
and homogeneous; they can be written in the form from the
previous section, ${\vec F} _{drag} = -\beta(p) {\vec p}$.  
(to simplify notation we henceforth drop the vector symbol over the $p$
denoting by $p$ both $|p|$ and ${\vec p}$, unambigious by context). 
In the limit of weak damping we expect $L$ to
be approximately constant so that, time averaging, we arrive at 
$<\delta L> = -<\beta(p)> L$
to leading order in $\beta(p)$. 
Note also that
$\delta U = -\beta(p) p^2$ to leading order in $\beta$.

Note that the time derivative of ${\cal R}$ is the sum of a Poisson 
bracket with H plus a term probprtional to $\beta$ (see \Eref{4}). 
This is, ${{{\rm d}{\cal R}}\over{{\rm d t}}} = 2{\cal V} + \cal{O}(\beta)$ which, 
averaged over bounded orbits, indicates 
(the virial theorem) that $<{\cal V}> =\cal{O}(\beta)$. 
Thus for $V(r) = kr^\alpha$ this implies that 
$<p^2> = \alpha <V>+{\cal O}(\beta)$ so that 
$<U> = {{\alpha+2}\over{2}} <V> + {\cal O}(\beta)$, 
which to leading order in $\beta$ in \Eref{11} indicates
\BE
-{{2\alpha}\over{<p^2>}} \biggl( <\beta(p)p^2> - <\beta(p)><p^2> \biggr) = 
{{f'}\over{f}} \delta(d/c)
\label{12}
\EE
Thus restricted to
linear damping 
($\beta$ constant), but for any  
$\alpha$, the averages in \eref{12} factorize trivially
and the aspect ratio 
is unchanged to leading order under linear damping. 
The \Eref{12} also indicates that
this will, in general, not be the case for 
a velocity dependent damping coefficient.  
Although for potentials with a 
fixed scaling exponent 
$\alpha$ there is but one dimensionless parameter 
(See LHS \Eref{10}), the introduction 
of the damping coefficient $\beta$ introduces new length and time scales,
indicating that the orbital aspect ratio $d/c$ may be a 
function of $\beta$ and time. 
The fact that $\beta$ is time odd does apparently not preclude
its inclusion to linear order in the orbital aspect ratio in general.
Thus, we repeat, the conclusion that for any monomial 
potentials linear damping preserves the orbital shape is not a
consequence of dimensional analysis and discrete symmetries. 

As a final check, note that
relation \Eref{12} and \Eref{26} are both 
consistent with the attractors of the secular flow in the orbital shape.  
For circular orbits $p^2$ is a constant of the motion (again to 
leading order in $\beta$) and thus the 
LHS of \Eref{12} is zero, as expected by symmetry. 
Note that in contrast to \eref{12} in \Eref{26} the change in the 
eccentricity is proportional to the eccentricity for any 
$\beta(p)$, and since the eccentricity vanishes in this limit its 
orbit-averaged change by \eref{26} does as well. 
Also the strictly radial infall orbit limit is one in which 
the inner radius, $d \rightarrow 0$,  and so the LHS of \Eref{12} 
being non-zero in this limit 
looks inconclusive. But, by the definition of $f$ via \Eref{10} we 
see that in this limit 
$f \rightarrow 0$ or $f \rightarrow \infty$ depending on the 
sign of $\alpha$. Thus, by \eref{9}, $L=0$ and remain zero for
any $\beta(p)$.
In Tarasov's formulation, since \eref{26} is fully vector covariant
for isotropic and homogeneous (but otherwise arbitrary $\beta(p)$) 
the change in the ${\vec S}$ must be along the vector itself
for the radial infall case. 
Furthermore, as indicated in the discussion following \eref{26}, the 
change $\Delta {\vec S}$ is linear in ${\vec L}$ (for any $\beta(p)$) 
which vanishes in the radial infall case. 
In summary, both prescriptions indicate 
that circular orbits and radial infall 
must satisfy $<\dot {\vec S}> = 0$ 
{\it for any damping function} as expected on the grounds by  
symmetry. 

Furthermore, it is straightforward to go further and perturbatively
show using \eref{12} and the equations of motion
that $\beta = const.$ is the {\it only} 
shape-preserving damping function for the Kepler potential ($\alpha = -1$).  
The result is clearly common to all monomial central potentials only, as it 
is straightforward to demonstrate a counterexample in a more complicated
potential. 
This is due to the fact that there are no additional length 
scales in the potential and is not the case with other
potentials, such as the effective potential in General Relativity
(where the Schwarzschild radius arises as a second length scale in 
the potential). 

Returning to the rather general statement \eref{26}, in the Tarasov
formulation, the explicit time dependence of a candidate constant of motion 
${\cal O}$ gives a second term which 
cancels the last two terms. If the operator has a 
fixed momentum scaling weight (for example, $L$ is weight 1 and $S$ is 
essentially wieght 2), the last two terms will be of that same scaling weight 
only for the case of linear friction, $\beta(p) = const.$ 
Note that this argument does not rule out the existence of 
additional constants of the motion in the dissipative system 
that scale to zero as one goes to the Hamiltonian limit. 
The argument does, however, certify that in the case of linear 
friction the original Hamiltonian symmetries do survive to leading order
in that friction.

%

Having shown that linear friction preserves the eccentricity to 
leading order begs the question of what happens in higher order in the damping.  In the 
spirit of the discussion after \eref{10} where the Virial played a 
key role, consider the time evolution of that part of the dynamical algebra 
\BE
{\dot {\cal R}} =  \{ {\cal R} , H\} -\beta(p) {\cal R}= 2{\cal V} - \beta(p)
{\cal R}
\label{4}
\EE

\BE 
\fl {\dot {\cal V}} = 
\{{\cal V} , H\} -2\beta(p)(2{\cal V} -H) 
= -{{1}\over{r}} (\partial_r V + 
{{1}\over{2}} \partial_r(r\partial_r V)) {\cal R} 
 -2\beta(p)(2{\cal V} -H) 
\label{5}
\EE

\BE
{\dot H} = -2\beta(p) (2{\cal V} - H) 
\label{5prime}
\EE
and for completeness, we have
\BE 
\{{\cal R} , {\cal V} \} = H +{\cal V} -V+{{r}\over{2}} \partial_r V
+ {{r}\over{2}}\partial_r(r\partial_r V)
\label{6}
\EE
To orient the reader to the content of these, first note the 
Hamiltonian limit ({\it i.e.} $\beta(p) \rightarrow 0$ limit) 
for the Kepler case ($\alpha = -1$), both
$<{\cal V}> \rightarrow 0$ and $<{\cal R}/r^3> \rightarrow 0$
as expected.  
We thus expect both of these time averages to be atleast proportional 
to some positive power of $\beta$. Now, in the abscence of damping
${\cal V}$ is time even and ${\cal R}$ is time odd. Formally, 
taking $\beta$ to be time odd preserves this discrete symmetry of the 
above evolution equations. Since we expect the $<{\cal V}>$ and $<{\cal R}>$ 
to be analytic functions of $\beta$, it must thus 
be that $<{\cal V}>$ vanishes 
quadratically as $\beta \rightarrow 0$. 

An elementary argument now certifies that the $<{\cal V}>$ must 
be nonpositive in the damped system. Take $\beta(p) = \beta$ a constant. 
Consider the radial component of the velocity, ${\cal R}/r$. 
It must average to zero in the $\beta \rightarrow 0$ limit. Since the 
damped orbit must shrink, we thus expect $<{\cal R}/r> \sim < -C\beta>$ for 
some positive quantity $C$ (a function 
of the other orbital parameters, etc.). But now take the evolution equation 
\eref{4} divide by $r$ and time average. Clearly, integrating 
by parts, $<{\dot {\cal R}}/r> = -<{\cal R}/r^2 ~ {\dot r}> 
= -<{\cal R}^2/r^3>$ implying that the time average  
$<2{\cal V}/r - \beta {\cal R}/r>$  
must also be strictly negative. But since  $<{\cal R}/r>$ must already 
be negative, the $<{\cal V}>$ must also be strictly 
negative in the damped system. 

Specializing to Kepler ($\alpha = -1$), differentiating 
$\epsilon^2 = |{\vec S}|^2$ in time and applying the 
equations of motion of the system with friction, we learn 
that $\frac{d}{dt}\epsilon^2  = -8\beta L^2 {\cal V}$. Using the fact that 
$<\cal V>$ is negative and order $\beta^2$ and integrating both sides, 
we learn that the asymptotic change in the eccentricity to leading order
is positive and also of order $\beta^2$ (note the integral 
itself scales as $1/\beta$). Furthermore, since these are exact evolution 
equations, we have shown that the integration is 
well behaved throughout. Thus linear friction causes Kepler 
orbits to become more eccentric by a fixed 
amount that scales with the square of the linear damping coefficient. 

\section{Conclusion} 
Typically Hamiltonian symmetries lose thier 
relevance to the geometry of the trajectories 
when damping forces are added to the Hamiltonian system. 
If the damping is weak, homogeneous and isotropic, then 
for linear damping in monomial potentials, we have shown 
that orbit-averaged shape is stationary. 
This can understood most easily through 
Tarasov's generalization of conserved 
quantities from the Hamiltonian context to the non-Hamiltonian setting. 
This approach also quantifies in precise analytic 
terms the fate and subsequent utility of the dynamical symmetry 
algebra in the associated non-Hamiltonian system. 

There are three main frameworks for understanding orbital motion in a 
perturbed central 
field. The first 
is directly from the equations of motion; this admits straightforward 
generalization to the 
non-Hamiltonian case but 
somewhat obscures the structure and fate of the dynamical 
symmetry group. 
The second, namely the KS construction, embeds the 
Kepler orbit 
problem in the higher dimensional set of 
harmonic oscillators with constraints; this illuminates the dynamical symmetry 
group but does not 
seem to readily admit a generalization to the non-Hamiltonian system. Lastly, 
the geometrical 
approach, namely that which associates the Kepler Hamilton equations to 
geodesic flow on manifolds
of constant curvature, also illuminates the dynamical symmetry 
group while making 
the generalization to the non-Hamiltonian case somewhat unclear. 

In light of these difficulties, 
we used Tarasov's framework (and applied to the damped central 
field problem here) for extending Poisson symmetries to dissipative 
systems, emphasising its utility
in making crisp connections between 
dynamics, algebra and the geometric character of the solutions. 
Finally, 
the dynamical algebra remains whole in first order in the linear 
dissipative system, but flow at higher order 
is not trivial. 
The secular perturbative 
method ``variation of constants'' is not 
adequate to explain this, however an elementary method based on
the Virial subalgebra explains the change in the shape of kepler 
orbits in higher order in linear damping.

\ack This work was supported in part by
the National Science Foundation through a grant to the Institute of
Theoretical Atomic and Molecular Physics at Harvard University and the
Smithsonian Astrophysical Observatory, where this work was begun and 
by a fellowship from the Radcliffe Institute for Advanced Studies where
this work was concluded. It is a pleasure to acknowledge useful discussions 
with Harvard-Smithsonian Center for Astrophysics
personell Mike Lecar, Hosein Sadgehpour, Thomas Pohl and Matt Holman. 


\vfill
\par 
\eject

\ \ 
\vskip .4in\

\end{document}